\documentclass{article}
\usepackage{amsmath,amssymb,color,url,amsthm}
\usepackage{multirow}
\allowdisplaybreaks
\def\theequation{\thesection.\arabic{equation}}
\def\d   {\displaystyle}
\newtheorem{thm}{Theorem}[section]

\newtheorem{algo}[thm]{Algorithm}
\newtheorem{ass}[thm]{Assumption}

\theoremstyle{definition}

\newtheorem{rem}{Remark}
\def\l     {\left}
\def\r     {\right}

\def\bbE   {{\mathbb E}}
\def\bbN   {{\mathbb N}}
\def\bbP   {{\mathbb P}}
\def\bbQ   {{\mathbb Q}}
\def\bbR   {{\mathbb R}}
\def\calF  {{\mathcal F}}
\def\tN    {\widetilde{N}}
\def\tf    {\widetilde{f}}
\def\tg    {\widetilde{g}}
\def\wtbeta {\widetilde{\beta}}
\def\olS   {\overline{S}}
\usepackage[dvipdfmx]{graphicx}
\usepackage{tikz}
\usetikzlibrary{positioning,intersections,calc,arrows.meta,math}
\usepackage{ascmac}
\begin{document}
\title{Monte Carlo simulation for Barndorff-Nielsen and Shephard model under change of measure}
\author{Takuji Arai\footnote{Corresponding author} \ \footnote{
Department of Economics, Keio University, 2-15-45 Mita, Minato-ku, Tokyo, 108-8345, Japan. \\ (arai@econ.keio.ac.jp)} \\
Yuto Imai\footnote{
Faculty of International Politics and Economics, Nishogakusha University, 6-16 Sanbancho, Chiyoda-ku, Tokyo, 102-8336, Japan. \\
(y-imai@nishogakusha-u.ac.jp)}}
\maketitle

\begin{abstract}
The Barndorff-Nielsen and Shephard model is a representative jump-type stochastic volatility model.
Still, no method exists to compute option prices numerically for the non-martingale case with infinite active jumps.
We develop two simulation methods for such a case under change of measure and conduct some numerical experiments. 
\end{abstract}

%
%
\setcounter{equation}{0}
\section{Introduction}
The Barndorff-Nielsen and Shephard (BNS) model, undertaken by Barndorff-Nielsen and Shephard \cite{BNS1} and \cite{BNS2},
is a non-Gaussian Ornstein-Uhlenbeck (OU) type stochastic volatility model and has been actively studied in recent years.
In option pricing for the BNS model, the Carr-Madan method based on the Fast Fourier Transform is well-known,
but it is available only when the discounted asset price process is a martingale.
Thus, this paper aims to develop simulation methods for the non-martingale BNS model under a martingale measure.

The asset price process $S$ in the BNS model is driven by a one-dimensional Brownian motion $W$ and a driftless subordinator $H_\lambda$,
where $\lambda>0$ is the time-scale parameter. 
The square of the volatility process $\sigma$ in the BNS model is given as an OU process driven by $H_\lambda$.
Since we can easily simulate $S$ for the case where $H_\lambda$ is finite active, we treat only the infinite active case.
In particular, we focus on the IG-OU type BNS model, a typical example of the BNS model with infinite active jumps.
Exact simulation methods for $\sigma^2$ in the IG-OU case have been developed by Qu et al. \cite{QDZ} and Sabino and Petroni \cite{SP}.
The algorithm proposed in \cite{SP} is used here. Its details can be found in Appendix.

On the other hand, when the discounted asset price process is not a martingale, we need to change the underlying probability measure,
denoted by $\bbP$, into an equivalent martingale measure in order to compute option prices.
Note that the uniqueness of equivalent martingale measures does not hold for the BNS model because the market is incomplete.
Therefore, we select the minimal martingale measure (MMM), one of the most representative equivalent martingale measures.
The MMM, denoted by $\bbP^*$, is an equivalent martingale measure that appears in discussions of local risk-minimization,
one of well-known optimal hedging strategies for incomplete markets. However, such background is not discussed here.
Note that the MMM is the only equivalent martingale measure that can be written down concretely among equivalent martingale measures
that appear in the context of hedging and risk management.
More importantly, $H_\lambda$ is no longer a L\'evy process under $\bbP^*$; more precisely,
$H_\lambda$ has neither independent nor stationary increments.
Thus, explicitly describing the characteristic function of $S$ under $\bbP^*$ is impossible, implying the Carr-Madan method is unavailable.

In light of the above, we develop two simulation algorithms for $S$ under $\bbP^*$, that is, the distribution of $S_T$ under $\bbP^*$,
where $T$ is the maturity of our market.
In the first algorithm, we define $Z$, the density process of $\bbP^*$,
as a martingale generated by the conditional expectation under $\bbP$ of the Radon-Nikodym density $d\bbP^*/d\bbP$
and compute the distributions of $Z_T$ and $S_T$ under $\bbP$ separately.
We make a seemingly rough approximation in the simulation of $Z$, but it performs unexpectedly accurately
when $\alpha$ the drift coefficient of $S$ is small. However, when $\alpha$ becomes large, this algorithm does not work.
On the other hand, we simulate $S$ under $\bbP^*$ directly in the second algorithm.
This algorithm is robust with respect to the value of $\alpha$.
The second algorithm relies on the acceptance/rejection (A/R) scheme.

This paper is organized as follows: We provide some mathematical preparations in the next section.
The two algorithms of our main objectives will be illustrated in Sections 3 and 4.
Section 5 is devoted to numerical experiments, and this paper concludes in Section 6.

%
%
\setcounter{equation}{0}
\section{Mathematical preliminaries}
Throughout this paper, we consider a financial market composed of one riskless asset and one risky asset.
We denote by $T>0$ the maturity of our market. Without loss of generality, we may assume that the interest rate is zero for simplicity.
Now, the risky asset price process $S=\{S_t\}_{0\leq t\leq T}$ is given as follows:
\begin{equation}\label{eq-BNS}
S_t:=S_0\exp\l\{\int_0^t\l(\mu-\frac{1}{2}\sigma_s^2\r)ds+\int_0^t\sigma_sdW_s+\rho H_{\lambda t}\r\}, \ \ \ t\in[0,T],
\end{equation}
where $S_0>0$, $\rho\leq0$, $\mu\in\bbR$, $\lambda>0$, $\sigma=\{\sigma_t\}_{0\leq t\leq T}$ is the volatility process defined below,
$W=\{W_t\}_{0\leq t\leq T}$ is a one dimensional standard Brownian motion, and
$H_\lambda=\{H_{\lambda t}\}_{0\leq t\leq T}$ is a subordinator without drift, that is, a driftless non-decreasing L\'evy process.
Let $N$ be the Poisson random measure of $H_\lambda$, and $\nu$ its L\'evy measure.
In addition, we define
\[
\tN(dt,dx):=N(dt,dx)-\nu(dx)dt,
\]
which is the so-called compensated Poisson random measure of $H_\lambda$.
Then, $S$ is a solution to the following stochastic differential equation (SDE):
\[
dS_t = S_{t-}\l(\alpha dt+\sigma_t dW_t + \int_0^\infty(e^{\rho x}-1)\tN(dt,dx)\r),
\]
where
\[
\alpha:=\mu+\int_0^\infty(e^{\rho x}-1)\nu(dx).
\]
Let $\{\calF_t\}_{0\leq t\leq T}$ be the filtration generated by $W$ and $H_\lambda$.
For more details on the BNS model, see Arai et al. \cite{AIS-BNS}, Nicolato and Venardos \cite{NV}, and Schoutens \cite{Scho}.

Now, we define the volatility process $\sigma$ as the square root of a solution $\sigma^2=\{\sigma^2_t\}_{0\leq t\leq T}$
to the following SDE: 
\begin{equation}\label{sigma-SDE}
d\sigma_t^2 = -\lambda\sigma_t^2dt+dH_{\lambda t}, \ \ \ \sigma_0^2>0.
\end{equation}
Here, the BNS model is said to be IG-OU type if the L\'evy measure $\nu$ is described as
\begin{equation}\label{eq-nu}
\nu(dx)=\frac{\lambda a}{2\sqrt{2\pi}}x^{-\frac{3}{2}}(1+b^2x)\exp\l\{-\frac{1}{2}b^2x\r\}dx, \ \ \ x\in(0,\infty),
\end{equation}
where $a,b>0$. In fact, $H_\lambda$ is infinite active since $\nu((0,\infty))=\infty$.
Throughout this paper, we consider the IG-OU type BNS model.

Next, we discuss the minimal martingale measure (MMM).
First of all, we denote by $M=\{M_t\}_{0\leq t\leq T}$ the martingale part of $S$. That is, it is described as
\[
dM_t=S_{t-}\l(\sigma_t dW_t + \int_0^\infty(e^{\rho x}-1)\tN(dt,dx)\r), \ \ \ M_0=0.
\]
An equivalent martingale measure $\bbP^*$ is called the MMM if any square integrable $\bbP$-martingale orthogonal to $M$ is also
a $\bbP^*$-martingale. Now, we define a martingale $Z=\{Z_t\}_{0\leq t\leq T}$ as
\[
Z_t:=\bbE\l[\frac{d\bbP^*}{d\bbP}\big|\calF_t\r],
\]
and call it the density process of $\bbP^*$. Note that $Z_0=1$ and $Z_T=d\bbP^*/d\bbP$.
Besides, $Z$ is a solution to the following SDE:
\begin{equation}\label{SDE-Z}
dZ_t=-Z_{t-}\Lambda_tdM_t,
\end{equation}
where
\[
\Lambda_t:=\frac{1}{S_{t-}}\frac{\alpha}{\sigma_t^2+C^\rho_2}
\]
and
\begin{equation}\label{eq-C2}
C^\rho_2:=\int_0^\infty(e^{\rho x}-1)^2\nu(dx)=2\rho\lambda a\l(\frac{1}{\sqrt{b^2-4\rho}}-\frac{1}{\sqrt{b^2-2\rho}}\r).
\end{equation}

\begin{ass}\label{ass}
Throughout this paper, we assume that
\begin{equation}\label{eq-ass}
\frac{b^2}{2}>2\l(\frac{1-e^{-\lambda T}}{\lambda}\vee|\rho|\r) \ \ \ \mbox{and} \ \ \ \frac{\alpha}{e^{-\lambda T}\sigma^2_0+C^\rho_2}>-1.
\end{equation}
The first and second conditions in (\ref{eq-ass}) ensure the martingale property of the product process $ZS$
and the positivity of the solution to the SDE (\ref{SDE-Z}), respectively. For more details on this matter, see \cite{AIS-BNS}.
\end{ass}

Solving (\ref{SDE-Z}) under Assumption \ref{ass}, we have
\begin{align}\label{eq-Z}
Z_t &= \exp\bigg\{-\int_0^tu_sdW_s-\frac{1}{2}\int_0^tu_s^2ds+\int_0^t\int_0^\infty\log(1-\theta_{s,x})\tN(ds,dx) \nonumber \\
    &  \hspace{5mm}+\int_0^t\int_0^\infty(\log(1-\theta_{s,x})+\theta_{s,x})\nu(dx)ds\bigg\},
\end{align}
where
\[
u_t:=\Lambda_tS_{t-}\sigma_t=\frac{\alpha\sigma_t}{\sigma_t^2+C^\rho_2} \ \ \ \mbox{ and } \ \ \ 
\theta_{t,x}:=\Lambda_tS_{t-}(e^{\rho x}-1)=\frac{\alpha(e^{\rho x}-1)}{\sigma_t^2+C^\rho_2}.
\]
Thus, the process $W^*=\{W^*_t\}_{0\leq t\leq T}$ defined by
\[
dW^*_t:=dW_t+u_tdt
\]
is a one-dimensional standard Brownian motion under $\bbP^*$. Similarly, we define
\begin{equation}\label{eq-nu*}
\nu^*_t(dx):=(1-\theta_{t,x})\nu(dx),
\end{equation}
which is corresponding to the L\'evy measure under $\bbP^*$, that is,
\[
\bbE_{\bbP^*}[N(dt,dx)|\calF_t]=\nu^*_t(dx)dt
\]
holds. Roughly speaking, defining
\[
\tN^*(dt,dx):=N(dt,dx)-\nu^*_t(dx)dt,
\]
we can see that stochastic integrals with respect to $\tN^*$ are a $\bbP^*$-martingale.
Note that $\nu^*_t(dx)$ depends on $t$ and has randomness. Thus, the Carr-Madan method is not available when $S$ is not a $\bbP$-martingale.
Using $W^*$, $\tN^*$ and $\nu^*_t$, we can describe $S$ as
\begin{align}\label{eq-BNS*}
S_t &= S_0\exp\bigg\{-\int_0^t\frac{1}{2}\sigma_s^2ds+\int_0^t\sigma_sdW^*_s \nonumber \\
    &\hspace{5mm} +\int_0^t\int_0^\infty\rho x\tN^*(ds,dx)+\int_0^t\int_0^\infty\l(\rho x+1-e^{\rho x}\r)\nu^*_s(dx)ds\bigg\}.
\end{align}

%
%
\setcounter{equation}{0}
\section{First Algorithm}
We illustrate an algorithm to compute the distributions of $Z_T$ and $S_T$ under $\bbP$ in order to simulate $S_T$ under $\bbP^*$.
We discretize the time interval $[0,T]$ into $M$ time steps and denote $\delta:=T/M$ and $t_k:=k\delta$ for $k=0,\dots,M$.
To simulate $Z_T$ and $S_T$, we compute sequentially $\log Z_{t_{k+1}}\big|_{\bbP}S_{t_k},\sigma^2_{t_k}$
and $\log S_{t_{k+1}}\big|_{\bbP}S_{t_k},\sigma^2_{t_k}$ for $k=0,\dots,M-1$,
where $X\big|_{\bbQ}Y_1,Y_2$ denotes the conditional distribution of $X$ under the probability measure $\bbQ$ given $Y_1$ and $Y_2$.

First, we discuss how to simulate $\log S_{t_{k+1}}$ given $S_{t_k}$ and $\sigma^2_{t_k}$. (\ref{eq-BNS}) implies that
\[
\log S_{t_{k+1}}
=\log S_{t_k}+\int_{t_k}^{t_{k+1}}\l(\mu-\frac{1}{2}\sigma_s^2\r)ds+\int_{t_k}^{t_{k+1}}\sigma_sdW_s+\rho \Delta H_{\lambda t_k},
\]
where $\Delta H_{\lambda t_k}:=H_{\lambda t_{k+1}}-H_{\lambda t_k}$.
From the view of the SDE (\ref{sigma-SDE}), we can approximate $\Delta H_{\lambda t_k}$ as
\[
\Delta H_{\lambda t_k}\approx \Delta\sigma^2_{t_k}+\lambda\sigma^2_{t_k}\delta,
\]
where $\Delta\sigma^2_{t_k}:=\sigma^2_{t_{k+1}}-\sigma^2_{t_k}$.
Note that we can compute $\sigma^2_{t_{k+1}}\big|_{\bbP}\sigma^2_{t_k}$ using Algorithm 1 in \cite{SP} described in Algorithm \ref{algo-SP}.
Thus, $\log S_{t_{k+1}}$ can be approximated as follows:
\begin{align*}
\log S_{t_{k+1}}
&\approx \log S_{t_k}+\l(\mu-\frac{1}{2}\sigma_{t_k}^2\r)\delta+\sigma_{t_k}\Delta W_{t_k}+\rho \Delta H_{\lambda t_k} \\
&\approx \log S_{t_k}+\l(\mu-\frac{1}{2}\sigma_{t_k}^2+\rho\lambda\sigma^2_{t_k}\r)\delta
         +\sigma_{t_k}\Delta W_{t_k}+\rho\Delta\sigma^2_{t_k},
\end{align*}
where $\Delta W_{t_k}:=W_{t_{k+1}}-W_{t_k}$.

Next, we mention an approximation of $\log Z_{t_{k+1}}$. By (\ref{eq-Z}), we have
\begin{align*}
\log Z_{t_{k+1}} &=\log Z_{t_k}-\int_{t_k}^{t_{k+1}}u_sdW_s+\int_{t_k}^{t_{k+1}}\int_0^\infty\log(1-\theta_{s,x})N(ds,dx) \\
                 &\hspace{5mm} +\int_{t_k}^{t_{k+1}}\l(-\frac{1}{2}u_s^2+\int_0^\infty\theta_{s,x}\nu(dx)\r)ds.
\end{align*}
Now, denoting
\[
K_t:=\frac{\alpha}{\sigma^2_t+C^\rho_2},
\]
we have $u_t=K_t\sigma_t$ and $\theta_{t,x}=K_t(e^{\rho x}-1)$. Thus, using the approximation
\begin{equation}\label{jump-approx}
\int_{t_k}^{t_{k+1}}\int_0^\infty\log(1-\theta_{s,x})N(ds,dx)\approx\log(1-\theta_{t_k,\Delta H_{\lambda t_k}}),
\end{equation}
we can approximate $\log Z_{t_{k+1}}$ as follows:
\begin{align*}
\log Z_{t_{k+1}} &\approx \log Z_{t_k}-K_{t_k}\sigma_{t_k}\Delta W_{t_k}+\log(1-\theta_{t_k,\Delta H_{\lambda t_k}}) \\
                 &\hspace{5mm} +\l(-\frac{1}{2}K_{t_k}^2\sigma^2_{t_k}+K_{t_k}C^\rho_1\r)\delta,
\end{align*}
where
\begin{equation}\label{eq-C1}
C^\rho_1:=\int_0^\infty(e^{\rho x}-1)\nu(dx)=\frac{\rho\lambda a}{\sqrt{b^2-2\rho}}.
\end{equation}
Remark that (\ref{jump-approx}) implements an approximation by combining all jumps occurring
in the time interval $[t_k,t_{k+1}]$ into a single jump.
This indicates that as the value of $K_{t_k}$ increases, i.e., as the value of $\alpha$ increases,
the accuracy worsens, as shown in the numerical experiments presented later.

The algorithm discussed here can be summarized as follows:

\begin{algo}\label{algo1}
\begin{enumerate}
\item Set the values of the parameters $S_0,\sigma_0,\lambda,a,b>0,\rho\leq0$, $\alpha\in\bbR$
      and $T>0$ so that satisfy the assumption (\ref{eq-ass}).
\item Set the values of the number of time steps $M\in\bbN$ and the number of paths $L\in\bbN$. Set $\delta=T/M$.
\item Compute $C^\rho_1$ and $C^\rho_2$ based on (\ref{eq-C1}) and (\ref{eq-C2}), respectively. Set $\mu=\alpha-C^\rho_1$.
\item For $l=1,\dots,L$, do
\begin{enumerate}
\item $S_{t_0,l}=S_0$, $\sigma_{t_0,l}=\sigma_0$ and $Z_{t_0,l}=1$.
\item For $k=0,\dots,M-1$, do
\begin{enumerate}
\item Compute $\sigma^2_{t_{k+1},l}$ by Algorithm \ref{algo-SP}.
\item Generate a random number $W\sim N(0,\delta)$.
\item $\d{\log S_{t_{k+1},l}=\log S_{t_k,l}+\sigma_{t_k,l}W+\rho\Delta\sigma^2_{t_k,l}}$ \\
      $\hspace{20mm}\d{+\l(\mu-\frac{1}{2}\sigma_{t_k,l}^2+\rho\lambda\sigma^2_{t_k,l}\r)\delta}$, \\
      where $\Delta \sigma^2_{t_k,l}=\sigma^2_{t_{k+1},l}-\sigma^2_{t_k,l}$.
\item Set $\d{K_{t_k,l}=\frac{\alpha}{\sigma^2_{t_k,l}+C^\rho_2}}$ and $\theta_{t_k,l}=K_{t_k,l}(e^{\rho\Delta H}-1)$,
      where $\d{\Delta H=\Delta \sigma^2_{t_k,l}+\lambda\sigma^2_{t_k,l}\delta}$.
\item $\d{\log Z_{t_{k+1},l}=\log Z_{t_k,l}-K_{t_k,l}\sigma_{t_k,l}W+\log(1-\theta_{t_k,l})}$ \\
      \hspace{20mm}$\d{+\l(-\frac{1}{2}K^2_{t_k,l}\sigma^2_{t_k,l}+K_{t_k,l}C^\rho_1\r)\delta}$.
\end{enumerate}
\item $S_{T,l}=S_{t_M,l}$ and $Z_{T,l}=Z_{t_M,l}$.
\end{enumerate}\end{enumerate}
\end{algo}

%
%
\setcounter{equation}{0}
\section{Second Algorithm}
This section develops another simulation method. Here we aim to compute $S_T$ under $\bbP^*$ directly.
As in the previous section, we discretize the time interval $[0,T]$ into $M$ time steps.
Unless otherwise noted, the notation defined in Sections 2 and 3 is also used here.

\subsection{Computation of $\sigma^2_{t_{k+1}}\big|_{\bbP^*}\sigma^2_{t_k}$}
We begin with the computation of $\sigma^2_{t_{k+1}}\big|_{\bbP^*}\sigma^2_{t_k}$.
From the view of Proposition 3.1 of \cite{QDZ}, (\ref{eq-nu}) and (\ref{eq-nu*}), we have
\begin{align}\label{eq-4-1}
\lefteqn{\bbE_{\bbP^*}\l[\exp\l\{-v\sigma^2_{t_{k+1}}\r\}\Big|\sigma^2_{t_k}\r]} \nonumber \\
&= \exp\l\{-vw\sigma^2_{t_k}\r\}\exp\l\{-\frac{1}{\lambda}\int_{vw}^v\frac{1}{u}\int_0^\infty(1-e^{-ux})\nu^*_{t_k}(dx)du\r\} \nonumber \\
&= \exp\l\{-vw\sigma^2_{t_k}\r\}\exp\l\{-\frac{1}{\lambda}\int_{vw}^v\frac{1}{u}\int_0^\infty(1-e^{-ux})\nu(dx)du\r\} \nonumber \\
&\hspace{5mm}\times\exp\l\{-K_{t_k}\frac{1}{\lambda}\int_{vw}^v\frac{1}{u}\int_0^\infty(1-e^{-ux})(1-e^{\rho x})\nu(dx)du\r\} \nonumber \\
&= \bbE\l[\exp\l\{-v\sigma^2_{t_{k+1}}\r\}\Big|\sigma^2_{t_k}\r] \nonumber \\
&\hspace{5mm}\times\exp\l\{-K_{t_k}\int_{vw}^v\frac{1}{u}\int_0^\infty(1-e^{-ux})(1-e^{\rho x})\frac{a\beta}{\sqrt{2\pi}}
   x^{-\frac{1}{2}}e^{-\beta x}dxdu\r\} \nonumber \\
&\hspace{5mm}\times\exp\l\{-K_{t_k}\int_{vw}^v\frac{1}{u}\int_0^\infty(1-e^{-ux})(1-e^{\rho x})\frac{a}{2\sqrt{2\pi}}
   x^{-\frac{3}{2}}e^{-\beta x}dxdu\r\}
\end{align}
for any $v>0$, where $w:=e^{-\lambda\delta}$, $\beta:=b^2/2$, and $\nu^*_t$ is decomposed into
\[
\nu^*_t(dx)=\l(1-K_t(e^{\rho x}-1)\r)\nu(dx)=\l(1+K_t(1-e^{\rho x})\r)\nu(dx).
\]
For $m=$1,3, converting $u$ and $x$ into $y=ux/v$ and $z=v/u$, we obtain
\begin{align*}
\lefteqn{\int_{vw}^v\frac{1}{u}\int_0^\infty(1-e^{-ux})(1-e^{\rho x})x^{-\frac{m}{2}}e^{-\beta x}dxdu} \\
&= \int_0^\infty(1-e^{-vy})\int_1^{\frac{1}{w}}(1-e^{\rho yz})(yz)^{-\frac{m}{2}}e^{-\beta yz}dzdy,
\end{align*}
and define a function $f_m$ and a constant $C_m$ as
\[
f_m(y):=\int_1^{\frac{1}{w}}(1-e^{\rho yz})(yz)^{-\frac{m}{2}}e^{-\beta yz}dz, \ \ \ y>0,
\]
and
\[
C_m:=\int_0^\infty f_m(y)dy=\frac{5-3m}{2}\sqrt{\pi}\lambda\delta\l(\beta^{\frac{m}{2}-1}-(\beta-\rho)^{\frac{m}{2}-1}\r).
\]
Since $f_m(y)>0$ for any $y>0$, $\tf_m(y):=f_m(y)/C_m$ gives a probability density function.
Consequently, we obtain
\begin{align*}
\lefteqn{\bbE_{\bbP^*}\l[\exp\l\{-v\sigma^2_{t_{k+1}}\r\}\bigg|\sigma^2_{t_k}\r]} \\
&= \bbE\l[\exp\l\{-v\sigma^2_{t_{k+1}}\r\}\bigg|\sigma^2_{t_k}\r]
   \exp\l\{-K_{t_k}\frac{a\beta}{\sqrt{2\pi}}\int_0^\infty(1-e^{-vy})C_1\tf_1(y)dy\r\} \\
&\hspace{5mm}\times\exp\l\{-K_{t_k}\frac{a}{2\sqrt{2\pi}}\int_0^\infty(1-e^{-vy})C_3\tf_3(y)dy\r\}
\end{align*}
for any $v>0$, which implies that $\sigma^2_{t_{k+1}}\big|_{\bbP^*}\sigma_{t_k}$ is given by
\begin{equation}\label{eq-4-2}
\sigma^2_{t_{k+1}}\big|_{\bbP}\sigma_{t_k}+\sum_{i=1}^{N_1}X_i^{(1)}+\sum_{i=1}^{N_3}X_i^{(3)},
\end{equation}
where $N_1$ and $N_3$ are random variables following the Poisson distribution with parameters $\d{K_{t_k}\frac{a\beta}{\sqrt{2\pi}}C_1}$
and $\d{K_{t_k}\frac{a}{2\sqrt{2\pi}}C_3}$, respectively,
and $\{X_i^{(1)}\}_{i\geq1}$ and $\{X_i^{(3)}\}_{i\geq1}$ are i.i.d. sequences with density functions $\tf_1$ and $\tf_3$, respectively.

\subsection{Computation of $X_i^{(m)}$}
In order to generate random variables $X_i^{(m)}$, $m=$1,3, we use the acceptance/rejection (A/R) scheme.
See Glasserman \cite{G} for details on the A/R scheme. Define
\[
g_1(y):=\frac{1}{C_1}y^{-\frac{1}{2}}\int_1^{\frac{1}{w}}z^{-\frac{1}{2}}dze^{-\beta y}
       =\frac{2}{C_1}y^{-\frac{1}{2}}e^{-\beta y}\l(\frac{1}{\sqrt{w}}-1\r), \ \ \ y>0.
\]
We have then $\tf_1(y)\leq g_1(y)$ for any $y>0$. Moreover, normalizing $g_1$ as
\[
\tg_1(y):=\frac{g_1(y)}{\int_0^\infty g_1(z)dz}=\sqrt{\frac{\beta}{\pi}}y^{-\frac{1}{2}}e^{-\beta y}, \ \ \ y>0,
\]
which is the density function of the Gamma distribution with shape parameter $1/2$ and scale parameter $1/\beta$.
Similarly, we define
\[
g_3(y):=\frac{|\rho|}{C_3}y^{-\frac{1}{2}}\int_1^{\frac{1}{w}}z^{-\frac{1}{2}}dze^{-\beta y}
       =\frac{2|\rho|}{C_3}y^{-\frac{1}{2}}e^{-\beta y}\l(\frac{1}{\sqrt{w}}-1\r), \ \ \ y>0.
\]
Since $1-e^{\rho yz}\leq|\rho|yz$, $\tf_3(y)\leq g_3(y)$ holds for any $y>0$,
and the normalization 
\[
\tg_3(y):=\frac{g_3(y)}{\int_0^\infty g_3(z)dz}, \ \ \ y>0,
\]
coincides with $\tg_1$. Under the above preparation, we describe an algorithm for computing $X_i^{(m)}$, $m=$1,3.

\begin{algo}\label{algo2-1}
\begin{enumerate}
\item Generate a random variable $Y$ following the Gamma distribution with shape parameter $1/2$ and scale parameter $1/\beta$.
\item Generate a random variable $U$ uniformly distributed on $[0,1]$.
\item If $U\leq\tf_m(Y)/g_m(Y)$ holds, then $X_i^{(m)}=Y$. Otherwise, reject $Y$ and go back to 1.
\end{enumerate}
\end{algo}

\begin{rem}
By Section 2.2.2 of \cite{G}, the probability that $Y$ is accepted in 3 of Algorithm \ref{algo2-1}, affecting computation time, is given by
\[
\frac{1}{\int_0^\infty g_m(z)dz}=\l\{
\begin{array}{ll}
\frac{\lambda\delta}{2\l(e^{\frac{1}{2}\lambda\delta}-1\r)}\frac{\sqrt{\beta-\rho}-\sqrt{\beta}}{\sqrt{\beta-\rho}}, & \mbox{ for }m=1, \\
  \frac{\lambda\delta}{|\rho|\l(e^{\frac{1}{2}\lambda\delta}-1\r)}\l(\sqrt{\beta(\beta-\rho)}-\beta\r),              & \mbox{ for }m=3.
\end{array}\r.
\]
\end{rem}

We further summarize the algorithm for computing $\sigma^2_{t_{k+1}}\big|_{\bbP^*}\sigma^2_{t_k}$ based on (\ref{eq-4-2}).
\begin{algo}\label{algo2-2}
\begin{enumerate}
\item Generate a random variable $N_1$ following the Poisson distribution with parameter $\d{K_{t_k}\frac{a\beta}{\sqrt{2\pi}}C_1}$.
\item If $N_1=0$, set $\Sigma_1=0$ and jump to 4.
      Otherwise, for $i=1,\dots,N_1$, generate a random variable $X_i^{(1)}$ by using Algorithm \ref{algo2-1}.
\item $\d{\Sigma_1=\sum_{i=1}^{N_1}X_i^{(1)}}$.
\item Generate a random variable $N_3$ following the Poisson distribution with parameter $\d{K_{t_k}\frac{a}{2\sqrt{2\pi}}C_3}$.
\item If $N_3=0$, set $\Sigma_3=0$ and jump to 7.
      Otherwise, for $i=1,\dots,N_3$, generate a random variable $X_i^{(3)}$ by using Algorithm \ref{algo2-1}.
\item $\d{\Sigma_3=\sum_{i=1}^{N_3}X_i^{(3)}}$.
\item Compute $\sigma^2_{t_{k+1}}\big|_{\bbP}\sigma_{t_k}$ by Algorithm \ref{algo-SP}.
\item Set $\sigma^2_{t_{k+1}}=\sigma^2_{t_{k+1}}\big|_{\bbP}\sigma_{t_k}+\Sigma_1+\Sigma_3$.
\end{enumerate}
\end{algo}

\subsection{Computation of $\log S_{t_{k+1}}\big|_{\bbP^*}S_{t_k}$}
By a similar argument with Section 3 and (\ref{eq-BNS*}), we approximate $\log S_{t_{k+1}}$ as follows:
\begin{align*}
\log S_{t_{k+1}}
&=       \log S_{t_k}-\int_{t_k}^{t_{k+1}}\frac{1}{2}\sigma_s^2ds+\int_{t_k}^{t_{k+1}}\sigma_sdW^*_s \\
&\hspace{5mm}+\int_{t_k}^{t_{k+1}}\int_0^\infty\rho xN(ds,dx)+\int_{t_k}^{t_{k+1}}\int_0^\infty\l(1-e^{\rho x}\r)\nu^*_{t_k}(dx)ds \\
&\approx \log S_{t_k}-\frac{1}{2}\sigma_{t_k}^2\delta+\sigma_{t_k}\Delta W^*_{t_k}
         +\rho \Delta H_{\lambda t_k}-C^\rho_1\delta+K_{t_k}C^\rho_2\delta, \\
&\approx \log S_{t_k}+\l(-\frac{1}{2}\sigma_{t_k}^2+\rho\lambda\sigma^2_{t_k}-C^\rho_1+K_{t_k}C^\rho_2\r)\delta
         +\sigma_{t_k}\Delta W^*_{t_k}+\rho\Delta\sigma^2_{t_k},
\end{align*}
where $\Delta W^*_{t_k}:=W^*_{t_{k+1}}-W^*_{t_k}$ and
\[
\Delta H_{\lambda t_k}:=H_{\lambda t_{k+1}}-H_{\lambda t_k}\approx\Delta\sigma^2_{t_k}+\lambda\sigma^2_{t_k}\delta.
\]
The algorithm for $\log S_{t_{k+1}}$ is described as follows:

\begin{algo}\label{algo2}
\begin{enumerate}
\item Conduct 1-3 in Algorithm \ref{algo1}.
\item For $l=1,\dots,L$, do
\begin{enumerate}
\item $S_{t_0,l}=S_0$ and $\sigma_{t_0,l}=\sigma_0$.
\item For $k=0,\dots,M-1$, do
\begin{enumerate}
\item $\d{K_{t_k,l}=\frac{\alpha}{\sigma^2_{t_k,l}+C^\rho_2}}$.
\item Generate a random number $W\sim N(0,\delta)$.
\item Compute $\sigma^2_{t_{k+1},l}$ by Algorithm \ref{algo2-2}.
\item $\d{\log S_{t_{k+1},l}
      =\log S_{t_k,l}+\l(-\frac{1}{2}\sigma_{t_k,l}^2+\rho\lambda\sigma^2_{t_k,l}-C^\rho_1+K_{t_k,l}C^\rho_2\r)\delta}$ \\
      $\d{+\sigma_{t_k,l}W+\rho(\sigma^2_{t_{k+1},l}-\sigma^2_{t_k,l})}$.
\end{enumerate}
\item $S_{T,l}=S_{t_M,l}$.
\end{enumerate}\end{enumerate}
\end{algo}

%
%
\setcounter{equation}{0}
\section{Numerical results}
We execute numerical experiments for Algorithms \ref{algo1} and \ref{algo2}.
Throughout our experiments, we use the parameter set calibrated in \cite{NV}, that is,
$S_0=$468.40, $\sigma^2_0=0.0041$, $\lambda=$2.4958, $a=$0.0872, $b=$11.98, $\rho=-4.7039$,
and vary the drift coefficient $\alpha$ from 0.01 to 100.
Furthermore, fix $T=1$ and set the number of time steps $M$ and the number of paths $L$ to 100 and 100,000, respectively.
The values of $M$ and $L$ are varied depending on the situation. Note that Assumption \ref{ass} is always satisfied in the above setting.
Here, we simulate the values of $\bbE_{\bbP^*}[S_T]$ and $\bbE_{\bbP^*}[\olS_T]$, where
\[
\olS_T:=\frac{1}{T}\int_0^TS_tdt,
\]
and confirm if the algorithms developed here are available to compute plain vanilla and Asian options.
As for $\bbE_{\bbP^*}[S_T]$, we compute
\[
\frac{1}{L}\sum_{l=1}^LS_{T,l}Z_{T,l} \ \ \ \mbox{ and } \ \ \ \frac{1}{L}\sum_{l=1}^LS_{T,l}
\]
by Algorithms \ref{algo1} and \ref{algo2}, respectively.
In addition, $\bbE_{\bbP^*}[\olS_T]$ is simulated computing
\[
\frac{1}{L(M+1)}\sum_{l=1}^L\sum_{k=0}^MS_{t_k,l}Z_{t_k,l} \ \ \ \mbox{ and } \ \ \ \frac{1}{L(M+1)}\sum_{l=1}^L\sum_{k=0}^MS_{t_k,l}
\]
by Algorithms \ref{algo1} and \ref{algo2}, respectively.
Since $S_0=\bbE_{\bbP^*}[S_T]=\bbE_{\bbP^*}[\olS_T]$, we define
\[
\mbox{Error}:=\frac{S_0-\mbox{simulation result}}{S_0}\times100
\]
and calculate it to evaluate the performance of our experiments.
Tables \ref{table1} and \ref{table2} give the results of Algorithms \ref{algo1} and \ref{algo2}, respectively.
Note that the top and bottom rows of the fourth columns in Tables \ref{table1} and \ref{table2} represent
the errors for the simulations of $\bbE_{\bbP^*}[S_T]$ and $\bbE_{\bbP^*}[\olS_T]$, respectively.
All the experiments have been conducted using Matlab R2022a with MacBook Air(2022) Apple M2 CPU, 24GB.

\begin{table}[htbp]\begin{center}
\begin{tabular}{lrrcr} \hline \vspace{-3.5mm} \\
 \multicolumn{1}{c}{$\alpha$} & \multicolumn{1}{c}{$M$} & \multicolumn{1}{c}{$L$}  & Error \%    & \multicolumn{1}{c}{Time sec.} \\ \hline \hline
 \multirow{2}{*}{0.01}        & \multirow{2}{*}{100}    & \multirow{2}{*}{100,000} & 0.012664767 & \multirow{2}{*}{30.785927}    \\
                              &                         &                          & 0.066203396 &                               \\ \hline
 \multirow{2}{*}{0.05}        & \multirow{2}{*}{100}    & \multirow{2}{*}{100,000} & 0.649362749 & \multirow{2}{*}{30.699075}    \\
                              &                         &                          & 0.802834253 &                               \\ \hline
 \multirow{2}{*}{0.1}         & \multirow{2}{*}{100}    & \multirow{2}{*}{100,000} & 0.795886643 & \multirow{2}{*}{30.954113}    \\
                              &                         &                          & 0.874496856 &                               \\ \hline
 \multirow{2}{*}{0.1}         & \multirow{2}{*}{500}    & \multirow{2}{*}{100,000} & 0.309957572 & \multirow{2}{*}{149.456563}   \\
                              &                         &                          & 0.339321885 &                               \\ \hline
 \multirow{2}{*}{1}           & \multirow{2}{*}{10,000} & \multirow{2}{*}{10,000}  & 95.28883837 & \multirow{2}{*}{349.566446}   \\
                              &                         &                          & 96.11253977 &                               \\ \hline
\end{tabular}\caption{Results of Algorithm \ref{algo1}.}\label{table1}\vspace{7mm}
\begin{tabular}{rrrcr} \hline \vspace{-3.5mm} \\
 \multicolumn{1}{c}{$\alpha$} & \multicolumn{1}{c}{$M$} & \multicolumn{1}{c}{$L$}  & Error \%     & \multicolumn{1}{c}{Time sec.} \\ \hline \hline
 \multirow{2}{*}{0.1}         & \multirow{2}{*}{100}    & \multirow{2}{*}{100,000} & -0.110181005 & \multirow{2}{*}{82.117931}    \\
                              &                         &                          & -0.059691138 &                               \\ \hline
 \multirow{2}{*}{1}           & \multirow{2}{*}{100}    & \multirow{2}{*}{100,000} & -0.467036359 & \multirow{2}{*}{185.662105}   \\
                              &                         &                          & -0.32856946  &                               \\ \hline
 \multirow{2}{*}{5}           & \multirow{2}{*}{100}    & \multirow{2}{*}{100,000} & -2.944984824 & \multirow{2}{*}{383.790356}   \\
                              &                         &                          & -2.125720502 &                               \\ \hline
 \multirow{2}{*}{5}           & \multirow{2}{*}{500}    & \multirow{2}{*}{100,000} & -0.482692379 & \multirow{2}{*}{604.576945}   \\
                              &                         &                          & -0.337696412 &                               \\ \hline
 \multirow{2}{*}{10}          & \multirow{2}{*}{100}    & \multirow{2}{*}{100,000} & -6.645643187 & \multirow{2}{*}{537.079512}   \\
                              &                         &                          & -4.989554592 &                               \\ \hline
 \multirow{2}{*}{10}          & \multirow{2}{*}{500}    & \multirow{2}{*}{100,000} & -1.163916967 & \multirow{2}{*}{740.789968}   \\
                              &                         &                          &-0.887464098  &                               \\ \hline
 \multirow{2}{*}{10}          & \multirow{2}{*}{1,000}  & \multirow{2}{*}{100,000} & -0.825495611 & \multirow{2}{*}{1042.507264}  \\
                              &                         &                          &-0.577846827  &                               \\ \hline
 \multirow{2}{*}{100}         & \multirow{2}{*}{20,000} & \multirow{2}{*}{10,000}  & -0.791441992 & \multirow{2}{*}{1341.982969}  \\
                              &                         &                          &-0.455007171  &                               \\ \hline
\end{tabular}\caption{Results of Algorithm \ref{algo2}.}\label{table2}\end{center}\vspace{-5mm}\end{table}

As for Algorithm \ref{algo1}, when $\alpha\leq$0.1, the errors are less than 1\%, which ensures sufficient accuracy.
As $\alpha$ increases, the accuracy becomes worse, but when $\alpha=$0.1, increasing $M$ to 500 improves accuracy.
Algorithm \ref{algo1} is accurate for small $\alpha$ despite including the rough approximation in (\ref{jump-approx}). 
However, when $\alpha=1$, the accuracy does not improve even if $M$ is increased.
In other words, it does not work for large $\alpha$.
On the other hand, Algorithm \ref{algo2} is sufficiently accurate for larger values of $\alpha$ if $M$ is increased.
For example, even in the extreme case of $\alpha=$100, the errors can be kept below 1\% taking $M=$20,000.
Note that when $M\geq$10,000, we reduce $L$ to 10,000 to save computation time.
However, it is a little bit more time-consuming than Algorithm \ref{algo1}. The use of the A/R scheme probably causes this.
Indeed, the acceptance probabilities in our experiments with $\delta=$0.01 are 0.0311 for $m=$1 and 0.978 for $m=3$; that is,
The A/R scheme for $m=$3 is very efficient but not efficient for $m=$1.
Finally, the above results show that Algorithms \ref{algo1} and \ref{algo2} are useful for calculating prices of
plain vanilla and Asian options.

%
%
\setcounter{equation}{0}
\section{Concluding remarks}
Two simulation algorithms for the non-martingale IG-OU type BNS model under the MMM have been developed in this paper.
Algorithm \ref{algo1} only works when the drift coefficient $\alpha$ is small, but it is faster than Algorithm \ref{algo2}.
On the other hand, Algorithm \ref{algo2} is sufficiently accurate by increasing $M$ the number of time steps for any value of $\alpha$.
Using the simulation algorithms developed here, option price calculations for the BNS model can also be implemented easily.
Hence, we can develop supervised deep learning to compute option prices by creating training samples with our simulation algorithms.
We left it to future work.

\setcounter{section}{0}
\renewcommand{\thesection}{\Alph{section}}
\setcounter{equation}{0}
\section{Computation of $\sigma^2_{t_{k+1}}\big|_{\bbP}\sigma_{t_k}$}
\renewcommand{\theequation}{A.\arabic{equation}}
We illustrate the algorithm for $\sigma^2_{t_{k+1}}\big|_{\bbP}\sigma_{t_k}$ developed by Sabino and Petroni \cite{SP}.
Suppose the values of parameters $\lambda,a,b>0$ are given.
Fix $T>0$ and $M\in\bbN$, and set $\delta=T/M$ and $t_k=k\delta$ for $k=0,\dots,M$.
Suppose that the value of $\sigma_{t_k}$ is given for some $k\in\{0,\dots,M-1\}$.
The following provides an algorithm for computing $\sigma^2_{t_{k+1}}$ under $\bbP$.

\begin{algo}[Algorithm 1 in \cite{SP}]\label{algo-SP}
\begin{enumerate}
\item Generate a random variable $X_1$ following the inverse Gaussian distribution with scale parameter $a(1-\sqrt{w})/b$
      and shape parameter $a^2(1-\sqrt{w})^2$, where $w=e^{-\lambda\delta}$.
\item Generate a random variable $N$ following the Poisson distribution with parameter $ab(1-\sqrt{w})$.
\item If $N=0$, set $X_2=0$ and jump to 5. Otherwise, for $n=1,\dots,N$, do
\begin{enumerate}
\item Generate a random variable $U$ distributed uniformly on $[0,1]$.
\item Set $V=\l(1+2\l(\frac{1}{\sqrt{w}}-1\r)U\r)^2$, and $\wtbeta=\beta V$.
\item Generate a random variable $J_n$ following the Gamma distribution with shape parameter $1/2$ and scale parameter $1/\wtbeta$.
\end{enumerate}\vspace{-3mm}
\item $\d{X_2=\sum_{n=1}^NJ_n}$.
\item $\sigma^2_{t_{k+1}}=w\sigma^2_{t_k}+X_1+X_2$.
\end{enumerate}
\end{algo}

\begin{center}
{\bf Acknowledgments}
\end{center}
Takuji Arai and Yuto Imai gratefully acknowledge the financial support of the MEXT Grant-in-Aid for
Scientific Research (C) No.22K03419 and Early-Career Scientists No. 21K13327, respectively.



\begin{thebibliography}{9999}
\bibitem{AIS-BNS} Arai, T., Imai, Y. and Suzuki, R. (2017). Local risk-minimization for Barndorff-Nielsen and Shephard models,
Finance \& Stochastics, 21 , pp.551-592.
\bibitem{BNS1} Barndorff-Nielsen, O. E., \& Shephard, N. (2001).
Modelling by L\'evy processes for financial econometrics. In L\'evy processes (pp.283-318). Birkh\"auser, Boston, MA.
\bibitem{BNS2} Barndorff-Nielsen, O. E., \& Shephard, N. (2001).
Non-Gaussian Ornstein-Uhlenbeck-based models and some of their uses in financial economics.
Journal of the Royal Statistical Society: Series B (Statistical Methodology), 63(2), pp.167-241.
\bibitem{G} Glasserman, P. (2004). Monte Carlo methods in financial engineering (Vol. 53, pp. xiv+-596). New York: Springer.
\bibitem{NV} Nicolato, E. and Venardos, E. (2003). Option pricing in stochastic volatility models of the Ornstein-\"Uhlenbeck type,
Mathematical Finance, 13 , pp.445-466.
\bibitem{QDZ} Qu, Y., Dassios, A., \& Zhao, H. (2021). Exact simulation of Ornstein-Uhlenbeck tempered stable processes.
Journal of Applied Probability, 58(2), pp.347-371.
\bibitem{SP} Sabino, P., \& Petroni, N. C. (2022).
Fast simulation of tempered stable Ornstein-\"Uhlenbeck processes. Computational Statistics, 37(5), pp.2517-2551.
\bibitem{Scho} Schoutens, W. (2003). L\'evy processes in finance: pricing financial derivatives, Wiley.
\end{thebibliography}
\end{document}